\documentclass[runningheads]{llncs}
\usepackage[T1]{fontenc}
\usepackage{graphicx}
\usepackage{bm}
\usepackage{amsmath}
\usepackage{amsfonts}
\usepackage{hyperref}
\usepackage{graphics}
\usepackage{tabularx}
\usepackage{booktabs}
\usepackage{color}
\usepackage{multirow}
\usepackage{siunitx}
\usepackage{appendix}
\usepackage{booktabs}

\begin{document}
\title{Adapting Frozen Mono-modal Backbones for Multi-modal Registration via Contrast-Agnostic Instance Optimization}
\titlerunning{Contrast-Agnostic Adaptation of Frozen Mono-modal Backbones for Multi-modal Registration}
%
\author{Yi Zhang
\and  Yidong Zhao 
\and Qian Tao
}
%
\authorrunning{Y. Zhang et al.}
\institute{Department of Imaging Physics, Delft University of Technology, The Netherlands}
\maketitle              

\begin{abstract}
Deformable image registration remains a central challenge in medical image analysis, particularly under multi-modal scenarios where intensity distributions vary significantly across scans. While deep learning methods provide efficient feed-forward predictions, they often fail to generalize robustly under distribution shifts at test time. A straightforward remedy is full network fine-tuning, yet for modern architectures such as Transformers or deep U-Nets, this adaptation is prohibitively expensive in both memory and runtime when operating in 3D. Meanwhile, the naive fine-tuning struggles more with potential degradation in performance in the existence of drastic domain shifts. In this work, we propose a registration framework that integrates a frozen pretrained \textbf{mono-modal} registration model with a lightweight adaptation pipeline for \textbf{multi-modal} image registration. Specifically, we employ style transfer based on contrast-agnostic representation generation and refinement modules to bridge modality and domain gaps with instance optimization at test time. This design is orthogonal to the choice of backbone mono-modal model, thus avoids the computational burden of full fine-tuning while retaining the flexibility to adapt to unseen domains. We evaluate our approach on the Learn2Reg 2025 LUMIR validation set and observe consistent improvements over the pretrained state-of-the-art mono-modal backbone. In particular, the method ranks second on the multi-modal subset, third on the out-of-domain subset, and achieves fourth place overall in Dice score. These results demonstrate that combining frozen mono-modal models with modality adaptation and lightweight instance optimization offers an effective and practical pathway toward robust multi-modal registration.
\begingroup
\renewcommand\thefootnote{}
\footnotetext{This paper was prepared as part of the Learn2Reg Challenge of MICCAI 2025.}

\addtocounter{footnote}{0}
\endgroup
\end{abstract}

%
%

\section{Introduction}
Medical image registration establishes anatomical correspondences between medical images and remains a prerequisite for applications such as treatment planning and longitudinal studies~\cite{staring2009registration,king2010registering,sotiras2013deformable}. Classical approaches optimize a similarity-regularization objective with respect to a parameterized transformation for each image pair~\cite{klein2007evaluation}. Recent advances in machine learning have reshaped medical image registration by replacing per-case optimization with data-driven prediction~\cite{rueckert2019model}. Unsupervised frameworks typically retain the classical objective which still combines an image similarity term with deformation regularization but optimize it through neural networks trained across many pairs in a population~\cite{balakrishnan2019voxelmorph,de2019deep}. Since the advent of U-Net~\cite{ronneberger2015unet}, a rich family of unsupervised registration architectures has emerged~\cite{balakrishnan2019voxelmorph,dalca2019unsupervised}. 
More recently, alternative backbones such as Transformers~\cite{chen2022transmorph}, implicit neural representations~\cite{wolterink2022implicit,van2023deformable}, and dual encoders~\cite{liu2024vector} have been explored to capture long-range dependencies and continuous deformation fields. 

A persistent challenge is \emph{cross-modality} registration, where varying MRI contrasts (\textit{e.g.}, T1w, T2w, FLAIR) disrupt simple intensity correspondences. This issue has been emphasized in the 2025 Learn2Reg LUMIR track, which evaluates robustness under contrast shifts and zero-shot settings~\cite{chen2025beyond,taha2023magnetic,marcus2007open,dufumier2022openbhb}. Classical cross-modality registration often employs mutual information (MI) and its variants~\cite{de2020mutual}, or modality-agnostic local descriptors such as MIND~\cite{heinrich2012mind}. Learning-based strategies include deep metric learning for multi-modal similarity~\cite{niethammer2019metric,mok2024modality}, image-to-image translation to synthesize target-like contrast~\cite{qin2019unsupervised}, and contrast-invariant training regimes~\cite{hoffmann2021synthmorph}. Despite these advances, modern pretrained backbones still suffer substantial performance drops when tested outside their native mono-modal domain.  

Instance optimization (IO) has emerged as a lightweight means of test-time adaptation by updating parameters for each image pair~\cite{balakrishnan2019voxelmorph,mok2023deformable,tian2024unigradicon}. However, full fine-tuning of large 3D models such as U-Nets or Transformers remains prohibitively expensive, and naive IO may be unstable under severe modality gaps. With the recent availability of foundational registration backbones~\cite{tian2024unigradicon,demir2024multigradicon}, the key question is how to adapt these powerful yet mono-modal models efficiently to multi-modal or out-of-domain scenarios.  

In this work, we introduce a contrast-agnostic instance optimization framework that acts as a general test-time adaptor for pretrained registration backbones. By taking the backbone output as initialization and applying a gated style-transfer module with lightweight refinement, the method remains orthogonal to backbone choice and incurs minor overhead. Beyond improving cross-modality robustness, it consistently enhances in-domain performance of state-of-the-art mono-modal baselines. Validated on the Learn2Reg 2025 LUMIR benchmark, our approach demonstrates competitive accuracy across in-domain, out-of-domain, and multi-modal tracks, establishing a lightweight and broadly applicable adaptor for robust multi-modal deformable registration.

\section{Methods}
\subsection{Deformable Image Registration}
Given a pair of 3D images $I_{\text{A}} \in \mathbb{R}^{D\times H\times W}$ and $I_{\text{B}} \in \mathbb{R}^{D\times H\times W}$, deformable registration seeks a dense transformation $\phi \in \mathbb{R}^{3 \times D\times H\times W}$ such that the warped source $I_{\text{A}} \circ \phi$ is anatomically aligned to the target $I_{\text{B}}$.  
As the deformation is typically small relative to the image grid $x$, it is expressed as $\phi(x) = x + u(x)$ with a displacement field $u$.  
The registration task can be formulated as the optimization
\begin{equation}
\hat{\phi} = \underset{\phi}{\operatorname{argmin}} \ 
\mathcal{L}_{\text{sim}}(I_{\text{A}} \circ \phi, I_{\text{B}}) + 
\lambda \mathcal{L}_{\text{reg}}(\phi),
\label{eq: phi optimization}
\end{equation}
where $\mathcal{L}_\text{sim}$ measures image similarity and $\mathcal{L}_\text{reg}$ regularizes the deformation field with trade-off parameter $\lambda > 0$.  

\begin{figure}[htbp]
    \centering
    \includegraphics[width=1\linewidth]{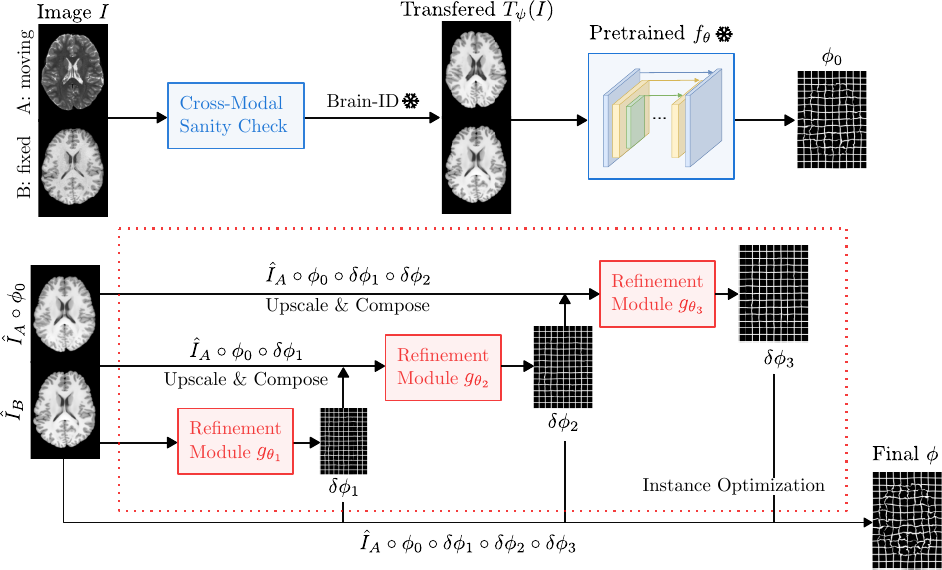}
    \caption{Overview of the proposed framework. Input images are optionally passed through a contrast-agnostic preprocessing (Brain-ID) when large modality differences are detected. A frozen pretrained backbone $f_{\theta}$ produces an initialization $\phi_0$, which is refined by instance optimization with lightweight multi-scale modules. The final deformation field $\phi_T$ is obtained by composing these refinements.}
    \label{fig:placeholder}
\end{figure}

\subsection{Instance Optimization Formulation}

Let $f_{\theta}$ denote the backbone network, pretrained on a mono-modal dataset. 
For each image pair $(I_A, I_B)$, the initial deformation field is estimated as 
\begin{equation}
    \phi_0 = f_{\theta}(I_A, I_B).
\end{equation}
However, due to distribution shifts, this initialization can be suboptimal. 
Several refinement strategies have been considered in the literature: 
(1) iterative application of $f_{\theta}$ by feeding the warped source $I_A \circ \phi_0$ as the new input; 
(2) direct optimization of the backbone parameters under the objective in Eq.~\ref{eq: phi optimization} \cite{tian2024unigradicon}; 
and (3) employing $\phi_0$ as an initialization for a subsequent registration stage, in analogy to classical sequential pipelines (e.g., affine initialization followed by deformable refinement).

Previous studies have reported that repeated application of a one-step network (e.g. VoxelMorph) leads to cumulative propagation of errors and poor accuracy~\cite{zhang2025bridging}. 
Moreover, strategy (2) entails prohibitively high computational requirements: a single forward–backward pass already exceeds 24 GB of VRAM for models with attention mechanism, rendering fine-tuning infeasible on widely accessible GPUs. 

Therefore, we adopt the third strategy: using $\phi_0$ and $I_A \circ \phi_0$ as initialization for a lightweight refinement stage. In this way, the pretrained backbone remains frozen, avoiding the cost of fine-tuning while still benefiting from its strong initialization. 
Since the refinement only acts on the output field, the approach is naturally orthogonal to the choice of backbone.

\subsection{Progressive Refinement Network}

Given the initial deformation $\phi_0 = f_{\theta}(I_A, I_B)$, 
we introduce a cascaded refinement module composed of three randomly initialized 3D U-Nets 
$\{g_{\theta_1}, g_{\theta_2}, g_{\theta_3}\}$. 
Each network predicts a residual deformation at progressively finer scales. 

At stage $t \in \{1,2,3\}$, the current warped source 
$I_A \circ \phi_{t-1}$ together with the target $I_B$ 
is passed to $g_{\theta_t}$ to estimate a residual field $\delta \phi_t$ at a reduced resolution $s_t$: 
\begin{equation}
    \delta \phi_t = g_{\theta_t}(I_A \circ \phi_{t-1}, I_B), 
    \quad s_t \in \left\{\tfrac{1}{4}, \tfrac{1}{2}, 1 \right\}.
\end{equation}
This residual is then upsampled to the original resolution and composed with the previous deformation:
\begin{equation}
    \phi_t = \phi_{t-1} \circ \mathrm{Upsample}(\delta \phi_t).
\end{equation}
Recursively, the full refinement process can be summarized as:
\begin{equation}
    \phi_T = \phi_0 \circ 
    \prod_{t=1}^{T} \mathrm{Upsample} \big(g_{\theta_t}(I_A \circ \phi_{t-1}, I_B)\big), 
    \quad T=3,
\end{equation}
where the product $\prod$ denotes successive composition of the deformation fields. 
The final field $\phi_T$ integrates both the pretrained initialization $\phi_0$ 
and the multi-scale residual refinements $\{\delta \phi_1, \delta \phi_2, \delta \phi_3\}$. The refinement admits a brief pretraining to provide more stable initialization, while still avoiding the computational burden of full fine-tuning.

\subsection{Contrast-agnostic Representation via Brain-ID}
When the input modality differs substantially from the training domain, the limited steps available for IO can lead to suboptimal adaptation given the inaccurate initialization. 
To mitigate such distribution gaps, we employ Brain-ID~\cite{liu2024brain}, a publicly available model that provides contrast-invariant representations and T1w-like style-transferred images. 
Given an input $I$, Brain-ID produces
\begin{equation}
    \hat{I} = T_{\psi}(I), \qquad z(I) = R_{\psi}(I),
\end{equation}
where $T_{\psi}$ is a frozen style-transfer module and $R_{\psi}$ is a contrast-robust encoder. 
We then use the transferred pair $(\hat{I}_A, \hat{I}_B)$ to compute the backbone initialization
\begin{equation}
    \phi_0 = f_{\theta}(\hat{I}_A, \hat{I}_B).
\end{equation}

Since both $T_{\psi}$ and $R_{\psi}$ are fixed pretrained modules trained on different data, they are not explicitly optimized for $f_{\theta}$. 
While Brain-ID improves robustness under distribution shifts, it may not preserve all T1w details. Therefore, we apply Brain-ID only when a fast (<0.5s in deployment) similarity check indicates a modality gap: if the low-resolution LNCC between $(I_A, I_B)$ falls below a threshold $\tau$, the pair is routed through Brain-ID; otherwise, the original images are used.

\subsection{Loss Functions}
We adopt LNCC with a Gaussian kernel of $9$ voxels as the similarity term $\mathcal{L}_{\text{sim}}$. 
At each refinement stage $t$, the warped source $I_A \circ \phi_t$ is compared with $I_B$, and the overall loss combines multi-level similarities with a diffusion regularizer on the final deformation $\phi_T$:
\begin{equation}
    \mathcal{L} \;=\; 
    \sum_{t=1}^{T} \mathcal{L}_{\text{sim}}(I_A \circ \phi_t, I_B)
    \;+\;
    \lambda \mathcal{L}_{\text{reg}}(\phi_T), 
    \quad T=3.
\end{equation}
When Brain-ID is used, $(I_A, I_B)$ are replaced by their transferred counterparts $(\hat{I}_A, \hat{I}_B)$.

\section{Experiments}
\subsection{Data}
We conducted our experiments on the 2025 L2R Challenge LUMIR dataset \cite{marcus2007open,dufumier2022openbhb,taha2023magnetic,chen2025beyond}, which consists of T1-weighted brain MRI scans from 10 public datasets. The dataset includes 3384 subjects for training, 36 subjects for validation which are available to the participants. All images were resampled and cropped to focus on the region of interest, resulting in an image size of
$[160,224,192]$. 

The validation set contains in-domain, out-of-domain, and multi-modal pairs, 
and evaluation is conducted on the official platform with Dice score, HD, non-diffeomorphic volume (NDV)~\cite{liu2024finite}, and target registration error (TRE) (on the in-domain subset).

\subsection{Experimental Settings}
We use the public weight of Vector Field Attention (VFA) backbone~\cite{liu2024vector} as $f_\theta$, 
the official baseline model for LUMIR 2025 challenge pretrained on mono-modal T1-weighted brain MRI. For Brain-ID transfer, we use a LNCC of window size of $11$ and $\tau = 0.4$ for modality detection. For the refinement modules, we use a standard U-Net for each module with a base channel counts of 32 and depth of 3 blocks. We use $\lambda = 0.1$ for Eq. \ref{eq: phi optimization} with a output magnitude scaling of 0.05 on $\phi_T$ in the finest resolution to avoid introducing large deformation. For IO, we use Adam optimizer, with a learning rate of 5e-4 and a linear warm up for 10 steps. The number of IO step is set to 50 steps in this work for the compatibility of finishing each pair of registraion within 1 mintue. We pretrained the refinement modules for 1000 steps on the Brain-ID transferred images on the training split for a more stable initialization with Adam optimizer and a learning rate of 1e-5. All experiments except official baselines on the leadboard are implemented on a workstation with NVIDIA RTX5090 with 32G of VRAM with PyTorch 2.8.0. All further implementation details will be included in the public code repository after the test phase evaluation.

We compared our method against both official challenge baselines and several related designs, including: 
(1) Zero displacement (raw), 
(2) ConvexAdam-MIND~\cite{siebert2024convexadam}, 
(3) SynthMorph~\cite{hoffmann2021synthmorph}, 
(4) VFA-SynthSR, where SynthSR~\cite{iglesias2023synthsr} normalizes inputs to T1w space before VFA, 
and (5) MultiGradICON (IO)~\cite{demir2024multigradicon} with 50-step IO. 

For our proposed pipeline, we also conducted ablation studies on our design choices, resulting in the following variants: 
(1) Brain-ID replacement (similar to VFA-SynthSR but replacing both inputs when non-T1w images are detected), 
(2) single U-Net instead of a multi-level cascade,
(3) cascaded refinement with additive updates instead of composition, 
and (4) cascaded refinement with weights initialized from 1000-step pretraining on Brain-ID.

In addition to the leaderboard submissions, we also examined several important baselines on the organizers’ reduced 10-class validation set due to the leaderboard submission limit. This includes: (1) VFA with multi-modal pretrained weight on brain MRI, provided by the official repository of LUMIR, (2) VFA-Brain-ID with full finetuning of the network parameters, (3) initializing refinement modules with different pretraining schedules (no pretrain, 1K steps, 5K steps). 
\section{Results and Discussion}

\begin{table}[hbtp]
\centering
\caption{Performance of official baselines and our variants on the validation leaderboard. 
Due to space constraints, TRE and HdDist95 are omitted; 
the full results are available on the official leaderboard.}
\scriptsize
\begin{tabular}{lccccc}
\toprule
Method & Overall DICE $\uparrow$ & NDV (\%) $\downarrow$ & In-domain $\uparrow$ & Out-of-domain $\uparrow$ & Multi-modal $\uparrow$ \\
\midrule
Zero Displacement & $0.5416 \pm 0.0341$ & - & 0.5668 & 0.5189 & 0.5391 \\
ConvexAdam-MIND~\cite{siebert2024convexadam} & $0.6732 \pm 0.0275$ & 0.0010 & 0.6998 & 0.6621 & 0.6578 \\
SynthMorph~\cite{hoffmann2021synthmorph} & $0.7012 \pm 0.0265$ & \textbf{0.0000} & 0.7262 & 0.6888 & 0.6888 \\
VFA-SynthSR~\cite{iglesias2023synthsr} & $0.7536 \pm 0.0260$ & 0.0074 & 0.7744 & 0.7545 & 0.7320 \\
MultiGradICON~\cite{demir2024multigradicon} & $0.7429 \pm 0.0236$ & 0.0017 & 0.7611 & 0.7453 & 0.7222 \\
\midrule
VFA-Brain-ID (no IO) & $0.7563 \pm 0.0239$ & 0.0791 & 0.7743 & 0.7548 & 0.7397 \\
VFA-Brain-ID (Single) & $0.7569 \pm 0.0237$ & 0.0675 & 0.7747 & 0.7555 & 0.7406 \\
VFA-Brain-ID (Add) & $0.7580 \pm 0.0241$ & 0.0539 & 0.7744 & 0.7593 & 0.7403 \\
VFA-Brain-ID (1K) & $\textbf{0.7601} \pm \textbf{0.0237}$ & 0.0896 & \textbf{0.7756} & \textbf{0.7617} & \textbf{0.7430} \\
\bottomrule
\label{tab:baseline}
\end{tabular}
\end{table}

\textbf{Main Comparison on Validation Leaderboard.} Table~\ref{tab:baseline} summarizes the performance of official baselines and our proposed variants on the public validation leaderboard. 
Compared to VFA-SynthSR, our Brain-ID based pipeline consistently achieves higher Dice on the multi-modal split, confirming the advantage of contrast-agnostic transfer in bridging modality gaps. 

The variants results in Table~\ref{tab:baseline} further highlight the importance of our architectural design. 
Multi-level cascaded refinement with compositional updates yields noticable gains over both the single-network and additive alternatives, while moderate pretraining (1K steps) provides the best overall Dice. 
These results suggest that lightweight refinement can effectively adapt frozen backbones without costly full-model tuning. In terms of efficiency, the entire refinement stage requires less than 10 GB GPU memory, making it substantially lighter than fine-tuning modern large 3D backbones.

\begin{table}[t]
\centering
\caption{Local validation results on the organizers’ reduced 10-class label set. 
Note that these results are for development only and may not perfectly correlate with the official leaderboard metrics.}
\scriptsize
\begin{tabular}{lcccc}
\toprule
Method & Overall DSC $\uparrow$ & In-domain $\uparrow$ & Out-of-domain $\uparrow$ & Multi-modal $\uparrow$ \\
\midrule
VFA Multi-modal (w/o IO) & $0.8874 \pm 0.0182$ & $0.9001 \pm 0.0085$ & $0.8828 \pm 0.0195$ & $0.8837 \pm 0.0187$ \\ 
VFA-Brain-ID (w/o IO) & $0.8896 \pm 0.0180$ & $0.9017 \pm 0.0088$ & $0.8896 \pm 0.0187$ & $0.8830 \pm 0.0190$ \\ 
VFA-Brain-ID (full finetune) & $0.8863 \pm 0.0188$ & $0.8990 \pm 0.0079$ & $0.8817 \pm 0.0204$ & $0.8820 \pm 0.0200$ \\
\midrule
VFA-Brain-ID (w/o pretrain) & $0.8930 \pm 0.0184$ & $0.9057 \pm 0.0086$ & $0.8941 \pm 0.0200$ & $0.8857 \pm 0.0181$ \\
VFA-Brain-ID (1K) & $\textbf{0.8938} \pm \textbf{0.0180}$ & $\textbf{0.9068} \pm \textbf{0.0072}$ & $\textbf{0.8945} \pm \textbf{0.0200}$ & $\textbf{0.8861} \pm \textbf{0.0178}$ \\
VFA-Brain-ID (5K) & $0.8929 \pm 0.0181$ & $0.9048 \pm 0.0074$ & $0.8933 \pm 0.0205$ & $0.8861 \pm 0.0180$ \\

\bottomrule
\end{tabular}
\label{tab:2}
\end{table}

\textbf{Additional Ablation Studies on Local Validation Subset.} Table~\ref{tab:2} reports additional ablations on the reduced 10-class validation set. 
The official multi-modal pretrained VFA achieves strong Dice on the multi-modal subset, but its in-domain and out-of-domain performance are inferior to our Brain-ID adaptor with similarity check, suggesting that contrast-agnostic test-time adaptation generalizes more reliably. 
Full finetuning of the VFA backbone performs worse than baseline without IO across all subsets, confirming that lightweight adaptation is more robust than exhaustive retraining of the whole model in this case. 
Finally, we observe that moderate pretraining of the refinement modules (1000 steps) yields the best overall Dice and out-of-domain accuracy, while longer adaptation (5000 steps) provides no additional gains and even slight degradation. 
These findings indicate that while limited pretraining can stabilize IO, excessive adaptation leads to diminishing returns. We emphasize that these findings are based on the reduced-label validation and should be considered indicative rather than definitive, 
as they may not fully correlate with the official leaderboard metrics.

\begin{figure}
    \centering
    \includegraphics[width=1\linewidth]{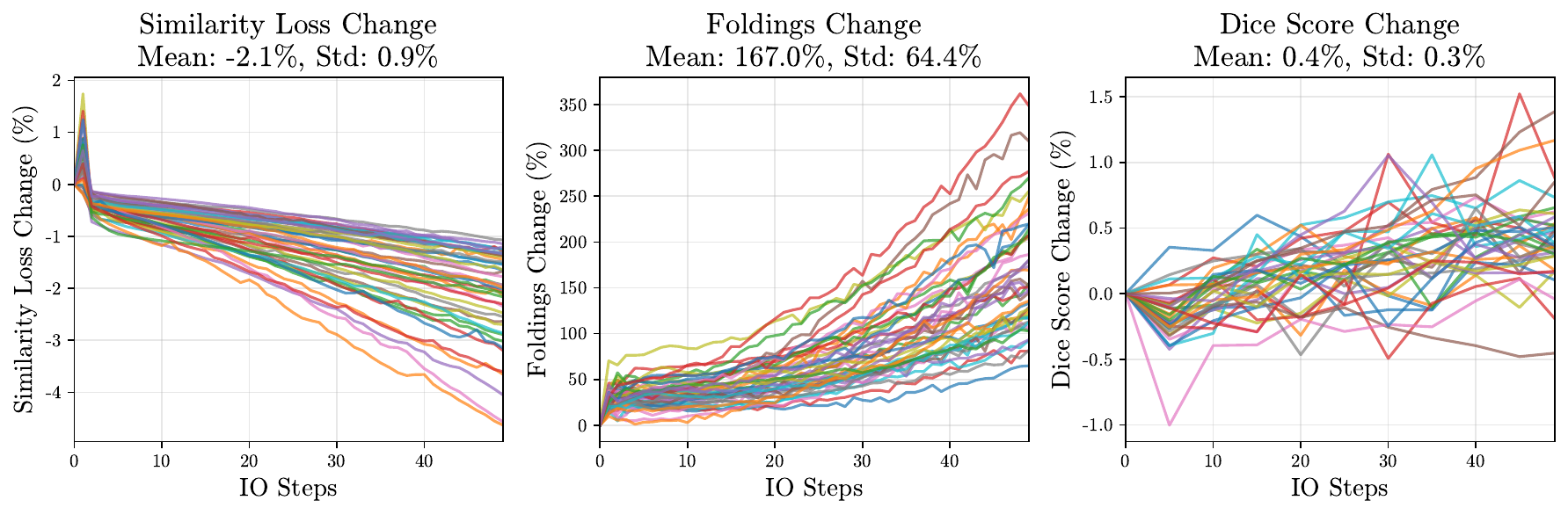}
    \caption{Dynamics of instance optimization on the 10-class validation subset with VFA-Brain-ID (1K). Dice score is recorded per 5 steps.}
    \label{fig:dynamics}
\end{figure}

\textbf{IO Dynamics \textit{w.r.t.} Losses and Dice Scores.} Fig.~\ref{fig:dynamics} illustrates the dynamics of instance optimization on the reduced 10-class validation set. 
On average, IO consistently improves both similarity loss and Dice, confirming that lightweight test-time adaptation can provide measurable gains even with frozen backbones. 
With a small regularization weight ($\lambda=0.1$), foldings increase slightly but remain mild given the low initial rate. 
Although case-wise Dice trajectories are variable and some instances exhibit small drops, such fluctuations are expected in a fully unsupervised setting and highlight opportunities for future refinement of IO strategies.

\section{Conclusion}
We introduced a flexible test-time adaptation framework that repurposes pretrained mono-modal backbones for robust multi-modal registration through contrast-agnostic transfer and lightweight instance optimization. 
The approach remains orthogonal to the backbone choice, avoids the prohibitive cost of full fine-tuning, and ranks among the top-performing methods on the Learn2Reg 2025 LUMIR benchmark.
Our ablation studies further confirm the benefit of Brain-ID based transfer, multi-level compositional refinement, and moderate pretraining. 
The results demonstrate that even frozen backbones can be effectively adapted to challenging cross-modality settings as a robust alternative to full network finetuning with the proposed framework.

\clearpage
\bibliographystyle{splncs04}
\bibliography{bib}

@article{zhang2025bridging,
  title={Bridging Classical and Learning-based Iterative Registration through Deep Equilibrium Models},
  author={Zhang, Yi and Zhao, Yidong and Tao, Qian},
  journal={arXiv preprint arXiv:2507.00582},
  year={2025}
}

@inproceedings{liu2024brain,
  title={Brain-id: Learning contrast-agnostic anatomical representations for brain imaging},
  author={Liu, Peirong and Puonti, Oula and Hu, Xiaoling and Alexander, Daniel C and Iglesias, Juan E},
  booktitle={European Conference on Computer Vision},
  pages={322--340},
  year={2024},
  organization={Springer}
}

@inproceedings{demir2024multigradicon,
  title={Multigradicon: A foundation model for multimodal medical image registration},
  author={Demir, Ba{\c{s}}ar and Tian, Lin and Greer, Hastings and Kwitt, Roland and Vialard, Fran{\c{c}}ois-Xavier and Est{\'e}par, Ra{\'u}l San Jos{\'e} and Bouix, Sylvain and Rushmore, Richard and Ebrahim, Ebrahim and Niethammer, Marc},
  booktitle={International Workshop on Biomedical Image Registration},
  pages={3--18},
  year={2024},
  organization={Springer}
}

@article{iglesias2023synthsr,
  title={SynthSR: A public AI tool to turn heterogeneous clinical brain scans into high-resolution T1-weighted images for 3D morphometry},
  author={Iglesias, Juan E and Billot, Benjamin and Balbastre, Ya{\"e}l and Magdamo, Colin and Arnold, Steven E and Das, Sudeshna and Edlow, Brian L and Alexander, Daniel C and Golland, Polina and Fischl, Bruce},
  journal={Science advances},
  volume={9},
  number={5},
  pages={eadd3607},
  year={2023},
  publisher={American Association for the Advancement of Science}
}

@article{siebert2024convexadam,
  title={Convexadam: Self-configuring dual-optimisation-based 3d multitask medical image registration},
  author={Siebert, Hanna and Gro{\ss}br{\"o}hmer, Christoph and Hansen, Lasse and Heinrich, Mattias P},
  journal={IEEE Transactions on Medical Imaging},
  year={2024},
  publisher={IEEE}
}

@inproceedings{niethammer2019metric,
  title={Metric learning for image registration},
  author={Niethammer, Marc and Kwitt, Roland and Vialard, Francois-Xavier},
  booktitle={Proceedings of the IEEE/CVF Conference on Computer Vision and Pattern Recognition},
  pages={8463--8472},
  year={2019}
}

@inproceedings{mok2024modality,
  title={Modality-agnostic structural image representation learning for deformable multi-modality medical image registration},
  author={Mok, Tony CW and Li, Zi and Bai, Yunhao and Zhang, Jianpeng and Liu, Wei and Zhou, Yan-Jie and Yan, Ke and Jin, Dakai and Shi, Yu and Yin, Xiaoli and others},
  booktitle={Proceedings of the IEEE/CVF Conference on Computer Vision and Pattern Recognition},
  pages={11215--11225},
  year={2024}
}

@article{heinrich2012mind,
  title={MIND: Modality independent neighbourhood descriptor for multi-modal deformable registration},
  author={Heinrich, Mattias P and Jenkinson, Mark and Bhushan, Manav and Matin, Tahreema and Gleeson, Fergus V and Brady, Michael and Schnabel, Julia A},
  journal={Medical image analysis},
  volume={16},
  number={7},
  pages={1423--1435},
  year={2012},
  publisher={Elsevier}
}

@inproceedings{qin2019unsupervised,
  title={Unsupervised deformable registration for multi-modal images via disentangled representations},
  author={Qin, Chen and Shi, Bibo and Liao, Rui and Mansi, Tommaso and Rueckert, Daniel and Kamen, Ali},
  booktitle={International Conference on Information Processing in Medical Imaging},
  pages={249--261},
  year={2019},
  organization={Springer}
}

@article{chen2025beyond,
  title={Beyond the LUMIR challenge: The pathway to foundational registration models},
  author={Chen, Junyu and Wei, Shuwen and Honkamaa, Joel and Marttinen, Pekka and Zhang, Hang and Liu, Min and Zhou, Yichao and Tan, Zuopeng and Wang, Zhuoyuan and Wang, Yi and others},
  journal={arXiv preprint arXiv:2505.24160},
  year={2025}
}

@article{liu2024vector,
  title={Vector field attention for deformable image registration},
  author={Liu, Yihao and Chen, Junyu and Zuo, Lianrui and Carass, Aaron and Prince, Jerry L},
  journal={Journal of Medical Imaging},
  volume={11},
  number={6},
  pages={064001--064001},
  year={2024},
  publisher={Society of Photo-Optical Instrumentation Engineers}
}

@article{balakrishnan2019voxelmorph,
	title        = {VoxelMorph: a learning framework for deformable medical image registration},
	author       = {Balakrishnan, Guha and Zhao, Amy and Sabuncu, Mert R and Guttag, John and Dalca, Adrian V},
	year         = 2019,
	journal      = {IEEE Transactions on Medical Imaging},
	publisher    = {IEEE},
	volume       = 38,
	number       = 8,
	pages        = {1788--1800}
}

@article{tian2024unigradicon,
  title={uniGradICON: A Foundation Model for Medical Image Registration},
  author={Tian, Lin and Greer, Hastings and Kwitt, Roland and Vialard, Francois-Xavier and Estepar, Raul San Jose and Bouix, Sylvain and Rushmore, Richard and Niethammer, Marc},
  journal={arXiv preprint arXiv:2403.05780},
  year={2024}
}

@article{klein2007evaluation,
	title        = {Evaluation of optimization methods for nonrigid medical image registration using mutual information and B-splines},
	author       = {Klein, Stefan and Staring, Marius and Pluim, Josien PW},
	year         = 2007,
	journal      = {IEEE Transactions on Image Processing},
	publisher    = {IEEE},
	volume       = 16,
	number       = 12,
	pages        = {2879--2890}
}

@inproceedings{mok2023deformable,
  title={Deformable medical image registration under distribution shifts with neural instance optimization},
  author={Mok, Tony CW and Li, Zi and Xia, Yingda and Yao, Jiawen and Zhang, Ling and Zhou, Jingren and Lu, Le},
  booktitle={International Workshop on Machine Learning in Medical Imaging},
  pages={126--136},
  year={2023},
  organization={Springer}
}

@article{dufumier2022openbhb,
  title={Openbhb: a large-scale multi-site brain mri data-set for age prediction and debiasing},
  author={Dufumier, Benoit and Grigis, Antoine and Victor, Julie and Ambroise, Corentin and Frouin, Vincent and Duchesnay, Edouard},
  journal={NeuroImage},
  volume={263},
  pages={119637},
  year={2022},
  publisher={Elsevier}
}

@article{taha2023magnetic,
  title={Magnetic resonance imaging datasets with anatomical fiducials for quality control and registration},
  author={Taha, Alaa and Gilmore, Greydon and Abbass, Mohamad and Kai, Jason and Kuehn, Tristan and Demarco, John and Gupta, Geetika and Zajner, Chris and Cao, Daniel and Chevalier, Ryan and others},
  journal={Scientific Data},
  volume={10},
  number={1},
  pages={449},
  year={2023},
  publisher={Nature Publishing Group UK London}
}

@article{king2010registering,
	title        = {Registering preprocedure volumetric images with intraprocedure 3-D ultrasound using an ultrasound imaging model},
	author       = {King, Andrew P and Rhode, Kawal S and Ma, Y and Yao, Cheng and Jansen, Christian and Razavi, Reza and Penney, Graeme P},
	year         = 2010,
	journal      = {IEEE Transactions on Medical Imaging},
	publisher    = {IEEE},
	volume       = 29,
	number       = 3,
	pages        = {924--937}
}

@article{staring2009registration,
	title        = {Registration of cervical MRI using multifeature mutual information},
	author       = {Staring, Marius and van der Heide, Uulke A and Klein, Stefan and Viergever, Max A and Pluim, Josien PW},
	year         = 2009,
	journal      = {IEEE Transactions on Medical Imaging},
	publisher    = {IEEE},
	volume       = 28,
	number       = 9,
	pages        = {1412--1421}
}

@article{sotiras2013deformable,
	title        = {Deformable medical image registration: A survey},
	author       = {Sotiras, Aristeidis and Davatzikos, Christos and Paragios, Nikos},
	year         = 2013,
	journal      = {IEEE Transactions on Medical Imaging},
	publisher    = {IEEE},
	volume       = 32,
	number       = 7,
	pages        = {1153--1190}
}

@article{rueckert2019model,
	title        = {Model-based and data-driven strategies in medical image computing},
	author       = {Rueckert, Daniel and Schnabel, Julia A},
	year         = 2019,
	journal      = {Proceedings of the IEEE},
	publisher    = {IEEE},
	volume       = 108,
	number       = 1,
	pages        = {110--124}
}

@article{de2019deep,
	title        = {A deep learning framework for unsupervised affine and deformable image registration},
	author       = {De Vos, Bob D and Berendsen, Floris F and Viergever, Max A and Sokooti, Hessam and Staring, Marius and I{\v{s}}gum, Ivana},
	year         = 2019,
	journal      = {Medical Image Analysis},
	publisher    = {Elsevier},
	volume       = 52,
	pages        = {128--143}
}

@inproceedings{de2020mutual,
	title        = {Mutual information for unsupervised deep learning image registration},
	author       = {de Vos, Bob D and van der Velden, Bas HM and Sander, J{\"o}rg and Gilhuijs, Kenneth GA and Staring, Marius and I{\v{s}}gum, Ivana},
	year         = 2020,
	booktitle    = {Medical Imaging 2020: Image Processing},
	volume       = 11313,
	pages        = {155--161},
	organization = {SPIE}
}

@article{hoffmann2021synthmorph,
	title        = {SynthMorph: learning contrast-invariant registration without acquired images},
	author       = {Hoffmann, Malte and Billot, Benjamin and Greve, Douglas N and Iglesias, Juan Eugenio and Fischl, Bruce and Dalca, Adrian V},
	year         = 2021,
	journal      = {IEEE Transactions on Medical Imaging},
	publisher    = {IEEE},
	volume       = 41,
	number       = 3,
	pages        = {543--558}
}

@inproceedings{ronneberger2015unet,
	title        = {U-net: Convolutional networks for biomedical image segmentation},
	author       = {Ronneberger, Olaf and Fischer, Philipp and Brox, Thomas},
	year         = 2015,
	booktitle    = {Medical Image Computing and Computer-Assisted Intervention--MICCAI 2015: 18th International Conference, Munich, Germany, October 5-9, 2015, Proceedings, Part III 18},
	pages        = {234--241},
	organization = {Springer}
}

@article{chen2022transmorph,
  title={Transmorph: Transformer for unsupervised medical image registration},
  author={Chen, Junyu and Frey, Eric C and He, Yufan and Segars, William P and Li, Ye and Du, Yong},
  journal={Medical image analysis},
  volume={82},
  pages={102615},
  year={2022},
  publisher={Elsevier}
}

@article{liu2024finite,
  title={On finite difference jacobian computation in deformable image registration},
  author={Liu, Yihao and Chen, Junyu and Wei, Shuwen and Carass, Aaron and Prince, Jerry},
  journal={International Journal of Computer Vision},
  pages={1--11},
  year={2024},
  publisher={Springer}
}

@article{marcus2007open,
	title        = {Open Access Series of Imaging Studies (OASIS): cross-sectional MRI data in young, middle aged, nondemented, and demented older adults},
	author       = {Marcus, Daniel S and Wang, Tracy H and Parker, Jamie and Csernansky, John G and Morris, John C and Buckner, Randy L},
	year         = 2007,
	journal      = {Journal of Cognitive Neuroscience},
	publisher    = {MIT Press One Rogers Street, Cambridge, MA 02142-1209, USA journals-info~…},
	volume       = 19,
	number       = 9,
	pages        = {1498--1507}
}

@article{dalca2019unsupervised,
	title        = {Unsupervised learning of probabilistic diffeomorphic registration for images and surfaces},
	author       = {Dalca, Adrian V and Balakrishnan, Guha and Guttag, John and Sabuncu, Mert R},
	year         = 2019,
	journal      = {Medical Image Analysis},
	publisher    = {Elsevier},
	volume       = 57,
	pages        = {226--236}
}

@inproceedings{wolterink2022implicit,
	title        = {Implicit neural representations for deformable image registration},
	author       = {Wolterink, Jelmer M and Zwienenberg, Jesse C and Brune, Christoph},
	year         = 2022,
	booktitle    = {Medical Imaging with Deep Learning},
	pages        = {1349--1359},
	organization = {PMLR}
}

@inproceedings{van2023deformable,
	title        = {Deformable Image Registration with Geometry-informed Implicit Neural Representations},
	author       = {van Harten, Louis and Van Herten, Rudolf Leonardus Mirjam and Stoker, Jaap and Isgum, Ivana},
	year         = 2023,
	booktitle    = {Medical Imaging with Deep Learning}
}
\end{document}